\documentclass[pra,aps,10pt,a4paper,nofootinbib,notitlepage,superscriptaddress,longbibliography]{revtex4-2}
\usepackage[utf8]{inputenc}
\usepackage[T1]{fontenc}
\usepackage{mathpazo}
\usepackage[scaled=0.86]{berasans}
\usepackage[scaled=1.03]{inconsolata}
\usepackage[UKenglish]{babel}
\usepackage[unicode,colorlinks]{hyperref}
\usepackage{amsmath,amssymb,amsthm}
\newcommand{\st}{\text{s.t.}}
\newtheorem{theorem}{Theorem}
\newcommand{\ket}[1]{\left| #1 \right\rangle}

\hypersetup{pdftitle={{Comment on "Geometry of the quantum set on no-signaling faces"}}}

\begin{document}
\title{Comment on ``Geometry of the quantum set on no-signaling faces''}
\author{Mateus Araújo}
\affiliation{Institute for Quantum Optics and Quantum Information (IQOQI), Austrian Academy of Sciences, Boltzmanngasse 3, 1090 Vienna, Austria}
\date{7th February 2023}

\begin{abstract}
In Ref.~\cite{rai2019} the authors claim that the Almost Quantum set of correlations cannot reproduce two points on the boundary of the quantum set of correlations. This claim is incorrect. The underlying issue is that the associated SDP is not strictly feasible, which makes the numerical solvers give unreliable answers. We give analytical proofs that both points are indeed reproduced by Almost Quantum.
\end{abstract}

\maketitle

In Ref.~\cite{rai2019} the authors study two lines on the set of correlations: the first is a convex combination between a PR-box $P$ and a local point in a ``quantum void'' $L$:
\begin{equation}\label{eq:l1}
l_1(\mu) = \mu P + (1-\mu) L,
\end{equation}
and the second is a convex combination between the same PR-box and a local point $H$ such that the line passes through the point corresponding to the correlations in the Hardy paradox \cite{Hardy1993}:
\begin{equation}\label{eq:l2}
l_2(\mu) = \mu P + (1-\mu) H.
\end{equation}
Here $P, L,$ and $H$ are given by
\begin{subequations}
\begin{equation}
P = \begin{pmatrix}
1 & 1/2 & 1/2 \\
1/2 & 1/2 & 1/2 \\
1/2 & 1/2 & 0
\end{pmatrix}
\end{equation}
\begin{equation}
L = \begin{pmatrix}
1 & 1/2 & 1/2 \\
1/2 & 1/3 & 1/3 \\
2/3 & 1/2 & 1/6
\end{pmatrix}
\end{equation}
\begin{equation}
H = \begin{pmatrix}
1 & \alpha & 2\alpha \\
\alpha & 0 & \alpha \\
2\alpha & \alpha & 0
\end{pmatrix}
\end{equation}
\end{subequations}
where $\alpha = \frac{9-\sqrt{5}}{38}$, and these numbers are tables of probabilities in Collins-Gisin notation, ordered as
\begin{equation}
\begin{pmatrix}
1 & p_B(0) & p_B(1) \\
p_A(0) & p(0,0) & p(0,1) \\
p_A(1) & p(1,0) & p(1,1)
\end{pmatrix}.
\end{equation}
The authors show that when $\mu$ is maximized under the constraints that $l_1(\mu)$ or $l_2(\mu)$ belong to the quantum set of correlations, the optimal values are $\mu_1^* = 0$ and $\mu_2^* = 5\sqrt{5}-11 \approx 0.1803$. They then claim that when the constraint is changed to the Almost Quantum set \cite{Navascues2015}, the optimal values become strictly larger than $\mu_1^*$ and $\mu_2^*$, which would imply that Almost Quantum cannot reproduce the boundary of the quantum set at these points.

The claim is based on numerical optimization over the Almost Quantum set using the NPA hierarchy \cite{Navascues2008}. However, when constraints on the probabilities are introduced, as is the case here, the NPA hierarchy is no longer necessarily strictly feasible, and a reliable numerical solution cannot be guaranteed. The authors used the solver SDPT3. We ran the SDP with the solvers SeDuMi and MOSEK and found the same numerical problems, which illustrates the fact that this is not an issue with a particular solver, but a fundamental one.

The issue is that numerical SDP solvers rely on primal-dual interior-point methods, that iteratively solve the primal and dual problems in order to approximate the optimal solution. They need primal and dual solutions corresponding to the optimal values to \emph{exist} in order to work reliably, and when strict feasibility fails this is not necessarily the case. Indeed, sufficient conditions for the existence of primal and dual optimal solutions are that the primal and dual problems are strictly feasible or, alternatively, the primal problem is strictly feasible and its set of feasible points is bounded \cite{NesterovBook}.

We'd like to emphasize that when an optimal dual solution does not exist, we cannot give an analytical proof of optimality based on SDP duality. We'll therefore reformulate the SDPs so that they admit optimal primal and dual solutions, which will both make the numerical solvers work reliably and allow an easy analytical proof.

The remainder of this Comment is organized as follows: in Section \ref{sec:trivial} we explain the technique for reformulating the SDPs and illustrate it via a trivial optimization problem, and in Section \ref{sec:proofs} we apply the technique to the problems of interest.

\section{Reformulating SDPs}\label{sec:trivial}

Consider the following primal-dual pair of SDPs:
\begin{equation}\label{primal}
\begin{aligned}
\min_X \quad & \langle C,X \rangle \\
\st \quad & \langle \Gamma_i, X \rangle = -b_i \quad \forall i,\\
& X \ge 0
\end{aligned}
\end{equation}
\begin{equation}\label{dual}
\begin{aligned}
\max_{y} \quad & \langle b, y \rangle \\
\st \quad & C + \sum_i y_i \Gamma_i  \ge 0
\end{aligned}
\end{equation}
We say that the primal problem \eqref{primal} is strictly feasible if there exists $X > 0$ such that $\langle \Gamma_i, X \rangle = -b_i$ for all $i$. The dual problem \eqref{dual} is strictly feasible if there exists $y$ such that $C + \sum_i y_i \Gamma_i  > 0$. Conversely, if the dual problem \eqref{dual} is \emph{not} strictly feasible, there exists a vector $\ket{v}$ such that for all feasible $y$ we have that $(C + \sum_i y_i \Gamma_i)\ket{v} = 0$. The analogous condition holds for the primal.

If one can find all such $\ket{v}$, one can then perform \emph{facial reduction} \cite{drusvyatskiy2017} on the SDP, reformulating it to lie on the orthogonal complement of these vectors, which will result in a smaller SDP that is necessarily strictly feasible. In fact, here we will not need to perform full facial reduction. Instead we will just use these implicit constraints $(C + \sum_i y_i \Gamma_i)\ket{v} = 0$ to simplify the SDP. It is easy to show that this simplified SDP admits optimal primal and dual solutions if and only if the one after full facial reduction does.

The crucial question is how to find these vectors $\ket{v}$. We'll use the Theorem of the Alternative for that \cite{drusvyatskiy2017}:
\begin{theorem}\label{thm:alternative}
Assume that the dual problem \eqref{dual} is feasible. Then exactly one of the following alternatives holds:
\begin{enumerate}
\item Problem \eqref{dual} is strictly feasible.
\item There exists $X \neq 0$ such that $X \ge 0,\ \langle C,X \rangle = 0$, and $\langle \Gamma_i, X \rangle = 0\ \forall i$.
\end{enumerate}
\end{theorem}
The vectors $\ket{v}$ will then be eigenvectors of the solutions of the second alternative with nonzero eigenvalue. Note that solving that problem is in general not easy. The approach we'll take here is to solve it numerically (as it is a feasibility SDP), and use the numerical solutions as guesses for the analytical solutions, which we can then easily verify. This approach prevents us from proving that we have found all solutions, but this is not a problem, since the final result does not hinge on that.

To illustrate the technique, let's consider now a trivial optimization problem where the lack of strict feasibility causes the same numerical issues. We want to maximize the CHSH functional $-p_A(0)-p_B(0) + p(0,0) + p(0,1) + p(1,0) - p(1,1)$ over the first level of the NPA hierarchy subject to the constraints $p_B(0) = p_B(1) = 0$. The resulting SDP, regarded most naturally as the dual form \eqref{dual}, is
\begin{equation}
\begin{gathered}
\max\quad  -p_A(0) + p(0,0) + p(0,1) + p(1,0) - p(1,1) \\
\st  \quad \begin{pmatrix}
1 & p_A(0) & p_A(1) & 0 & 0 \\
  & p_A(0) & a_{01} & p(0,0) & p(0,1) \\
  &        & p_A(1) & p(1,0) & p(1,1) \\
  &			 &	    &0       & b_{01} \\
  & 		&		&		& 0
\end{pmatrix} \ge 0
\end{gathered}
\end{equation}
The problem can be easily solved by hand, and the optimal value is 0. Solving it numerically with SeDuMi we get instead $1.5619\times 10^{-5}$ as an answer, and the information that the solver ran into numerical problems. To see why, consider the dual problem:
\begin{equation}
\begin{gathered}
\min\quad  \gamma_{00} \\
\st  \quad \begin{pmatrix}
\gamma_{00} & \gamma_{01} & \gamma_{02} & \gamma_{03} & \gamma_{04} \\
  & 1-2\gamma_{01} & 0 & -1/2 & -1/2 \\
  &        & -2\gamma_{02} & -1/2 & 1/2 \\
  &			 &	    &\gamma_{33}       & 0 \\
  & 		&		&		& \gamma_{44}
\end{pmatrix} \ge 0
\end{gathered}
\end{equation}
If we substitute the optimal value of the primal into the dual, $\gamma_{00} = 0$, positive semidefiniteness implies that $\gamma_{02} = 0$, which implies that the submatrix $\begin{pmatrix} -2\gamma_{02} & -1/2 \\ & \gamma_{33} \end{pmatrix}$ will necessarily have a negative eigenvalue, independently of the value of $\gamma_{33}$. Therefore there is no value of the $\gamma_{ij}$ that will make the dual solution match the optimal value of the primal. A closer examination reveals that as $\gamma_{00}$ gets smaller, $\gamma_{33}$ and $\gamma_{44}$ must get bigger and bigger, with the optimal value only being attained in the infinite limit\footnote{This implies that this SDP satisfies strong duality, which illustrates the fact that strong duality is not sufficient for a numerical solution to be found reliably.}. It's therefore unsurprising that a numerical solver cannot produce it.

Let's now reformulate the SDP to remove this problem. We could use the Theorem of the Alternative \ref{thm:alternative} to find the null eigenvectors, but the problem is so simple that the solution is immediately apparent: the zeros in the diagonal imply that the entire rows and columns where they are must be zero, so all feasible solutions must have the vectors $(0,0,0,1,0)^T$ and $(0,0,0,0,1)^T$ as null eigenvectors. Therefore the implicit constraints we must add are $p(0,0) = p(0,1) = p(1,0) = p(1,1) = b_{01} = 0$. The simplified SDP is therefore
\begin{equation}
\begin{gathered}
\max\quad  -p_A(0) \\
\st  \quad \begin{pmatrix}
1 & p_A(0) & p_A(1) & 0 & 0 \\
  & p_A(0) & a_{01} & 0 & 0 \\
  &        & p_A(1) & 0 & 0 \\
  &			 &	    &0  & 0 \\
  & 		&		&		& 0
\end{pmatrix} \ge 0
\end{gathered}
\end{equation}
which can now be solved numerically without issue, and its dual is
\begin{equation}
\begin{gathered}
\min\quad  \gamma_{00} \\
\st  \quad \begin{pmatrix}
\gamma_{00} & \gamma_{01} & \gamma_{02} & \gamma_{03} & \gamma_{04} \\
  & 1-2\gamma_{01} & 0 & \gamma_{13} & \gamma_{14} \\
  &        & -2\gamma_{02} & \gamma_{23} & \gamma_{24} \\
  &			 &	    &\gamma_{33}       & \gamma_{34} \\
  & 		&		&		& \gamma_{44}
\end{pmatrix} \ge 0
\end{gathered}
\end{equation}
which admits $\gamma_{ij} = 0 \ \forall i,j$ as a feasible solution that matches the primal optimal value.
\section{Proofs}\label{sec:proofs}

For the convenience of the reader, we have included two Mathematica notebooks reproducing the calculations that follow as ancillary files.

\subsection{First problem}

Let's now turn to the problem of maximizing $\mu$ under the constraint that $l_1(\mu)$, defined in Equation \eqref{eq:l1}, belongs to Almost Quantum. The resulting SDP, considered again as the dual form \eqref{dual}, is
\begin{equation}\label{eq:sdp1raw}
\begin{gathered}
\max\quad  \mu \\
\st  \quad \begin{pmatrix}
 1 & \frac{1}{2} & \frac{4-\mu }{6} & \frac{1}{2} & \frac{1}{2} & \frac{\mu +2}{6} & \frac{\mu +2}{6} & \frac{1}{2} & \frac{1-\mu }{6} \\
 & \frac{1}{2} & a_{01} & \frac{\mu +2}{6} & \frac{\mu +2}{6} & \frac{\mu +2}{6} & \frac{\mu +2}{6} & c_{01,0} & c_{01,1} \\
& & \frac{4-\mu }{6} & \frac{1}{2} & \frac{1-\mu }{6} & c_{01,0} & c_{01,1} & \frac{1}{2} & \frac{1-\mu }{6} \\
 & & & \frac{1}{2} &b_{01} & \frac{\mu +2}{6} & c_{0,01} & \frac{1}{2} & c_{1,01} \\
& & & & \frac{1}{2} & c_{0,01} & \frac{\mu +2}{6} & c_{1,01} & \frac{1-\mu }{6} \\
& & & & & \frac{\mu +2}{6} & c_{0,01} & c_{01,0} & c_{01,01} \\
& & & & & & \frac{\mu +2}{6} & c_{01,10} & c_{01,1} \\
& & & & & & & \frac{1}{2} & c_{1,01} \\
 &  & &  &  &  &  & & \frac{1-\mu }{6} \\
\end{pmatrix} \ge 0
\end{gathered}
\end{equation}
We solved numerically the SDP in the Theorem of the Alternative \ref{thm:alternative}, from which we guessed the solutions for $\ket{v}$ to be $(1,0,-1,0,-1,0,0,0,1)^T$ and $(0,0,0,1,0,0,0,-1,0)^T$, which we can verify analytically to be the case. The condition that they are null eigenvectors of the matrix in SDP \eqref{eq:sdp1raw} results in the following implicit equality constraints:
\begin{gather}
c_{1,01} = b_{01} \\
c_{01,01} = c_{0,01} \\
c_{01,10} = c_{0,01} \\
c_{01,0} = (\mu + 2)/6 \\
c_{01,1} = a_{01} - (1-\mu)/6 
\end{gather}
Substituting them into SDP \eqref{eq:sdp1raw}, we obtain the simplified SDP
\begin{equation}\label{eq:sdp1simple}
\begin{gathered}
\max\quad  \mu \\
\st  \quad \begin{pmatrix}
 1 & \frac{1}{2} & \frac{4-\mu }{6} & \frac{1}{2} & \frac{1}{2} & \frac{\mu +2}{6} & \frac{\mu +2}{6} & \frac{1}{2} & \frac{1-\mu }{6} \\
  & \frac{1}{2} & a_{01} & \frac{\mu +2}{6} & \frac{\mu +2}{6} & \frac{\mu +2}{6} & \frac{\mu +2}{6} & \frac{\mu +2}{6} & a_{01} - \frac{1-\mu}{6}  \\
  &  & \frac{4-\mu }{6} & \frac{1}{2} & \frac{1-\mu }{6} & \frac{\mu +2}{6} & a_{01} - \frac{1-\mu}{6}  & \frac{1}{2} & \frac{1-\mu }{6} \\
  &  &  & \frac{1}{2} & b_{01} & \frac{\mu +2}{6} & c_{0,01} & \frac{1}{2} & b_{01} \\
  &  &  &  & \frac{1}{2} & c_{0,01} & \frac{\mu +2}{6} & b_{01} & \frac{1-\mu }{6} \\
  &  &  &  &  & \frac{\mu +2}{6} & c_{0,01} & \frac{\mu +2}{6} & c_{0,01} \\
  &  &  &  &  &  & \frac{\mu +2}{6} & c_{0,01} & a_{01} - \frac{1-\mu}{6}  \\
  &  &  &  &  &  &  & \frac{1}{2} & b_{01} \\
  &  &  &  &  &  &  &  & \frac{1-\mu }{6} \\
\end{pmatrix} \ge 0
\end{gathered}
\end{equation}
which can now be solved numerically reliably, as there exists an optimal dual solution that matches the optimal value of the primal. From the numerical solution of the primal, we guess the values $\mu = 0, a_{01} = 1/3, b_{01} = 1/6$, and $c_{0,01} = 1/6$, which we can verify with a computer algebra system indeed provide a feasible solution for the primal.

We now solve numerically the dual of SDP \eqref{eq:sdp1simple}, and from the numerical solution guess the analytical solution
\begin{equation}
X^* = \frac12\begin{pmatrix}
 1 & -1 & -1 & 0 & -1 & 1 & 1 & 0 & 0 \\
  & 4 & 1 & 0 & 1 & -4 & -4 & 0 & 3 \\
  &  & 1 & 0 & 1 & -1 & -1 & 0 & 0 \\
  &  &  & 0 & 0 & 0 & 0 & 0 & 0 \\
  &  &  &  & 1 & -1 & -1 & 0 & 0 \\
  &  &  &  &  & 4 & 4 & 0 & -3 \\
  &  &  &  &  &  & 4 & 0 & -3 \\
  &  &  &  &  &  &  & 0 & 0 \\
  &  &  &  &  &  &  &  & 3 \\
\end{pmatrix}
\end{equation}
With a computer algebra system we very that $X^* \ge 0$ and that it satisfies the linear constraints, so it is indeed a feasible solution for the dual. Moreover, the dual objective at $X^*$ is 0, proving that 0 is indeed the optimal value of $\mu$.

\subsection{Second problem}
Let's now turn to the problem of maximizing $\mu$ under the constraint that $l_2(\mu)$, defined in Equation \eqref{eq:l2}, belongs to Almost Quantum. The resulting SDP, considered again as the dual form \eqref{dual}, is
\begin{equation}\label{eq:sdp2raw}
\begin{gathered}
\max\quad  \mu \\
\st  \quad \begin{pmatrix}
 1 & \frac{\mu }{2} + \alpha  (1-\mu ) & \frac{\mu }{2}+2\alpha(1-\mu) & \frac{\mu }{2} + \alpha  (1-\mu ) & \frac{\mu }{2}+2\alpha(1-\mu) & \frac{\mu }{2} & \frac{\mu }{2} + \alpha  (1-\mu ) & \frac{\mu }{2} + \alpha  (1-\mu ) & 0 \\
  & \frac{\mu }{2} + \alpha  (1-\mu ) & a_{01} & \frac{\mu }{2} & \frac{\mu }{2} + \alpha  (1-\mu ) & \frac{\mu }{2} & \frac{\mu }{2} + \alpha  (1-\mu ) & c_{01,0} & c_{01,1} \\
  &  & \frac{\mu }{2}+2\alpha(1-\mu) & \frac{\mu }{2} + \alpha  (1-\mu ) & 0 & c_{01,0} & c_{01,1} & \frac{\mu }{2} + \alpha  (1-\mu ) & 0 \\
  &  &  & \frac{\mu }{2} + \alpha  (1-\mu ) & b_{01} & \frac{\mu }{2} & c_{0,01} & \frac{\mu }{2} + \alpha  (1-\mu ) & c_{1,01} \\
  &  &  &  & \frac{\mu }{2}+2\alpha(1-\mu) & c_{0,01} & \frac{\mu }{2} + \alpha  (1-\mu ) & c_{1,01} & 0 \\
  &  &  &  &  & \frac{\mu }{2} & c_{0,01} & c_{01,0} & c_{01,01} \\
  &  &  &  &  &  & \frac{\mu }{2} + \alpha  (1-\mu ) & c_{01,10} & c_{01,1} \\
  &  &  &  &  &  &  & \frac{\mu }{2} + \alpha  (1-\mu ) & c_{1,01} \\
  &  &  &  &  &  &  &  & 0 \\
\end{pmatrix} \ge 0
\end{gathered}
\end{equation}
We immediately note that the 0 in the diagonal of the matrix implies that all variables in the last column must be zero, that is, $(0,0,0,0,0,0,0,0,1)^T$ must be a null eigenvector. From the numerical solution of the SDP in the Theorem of the Alternative \ref{thm:alternative} we find the other null eigenvectors, $(0,0,0,1,0,0,0,-1,0)^T$ and $(0,1,0,0,0,0,-1,0,0)^T$. Together they result in the implicit constraints
\begin{gather}
a_{01} = b_{01} = c_{01,1} = c_{1,01} = c_{01,01} = 0 \\
c_{01,0} = c_{0,01} = c_{01,10} = \mu/2.
\end{gather}
Substituting them into SDP \eqref{eq:sdp2raw}, we obtain the simplified SDP
\begin{equation}\label{eq:sdp2simple}
\begin{gathered}
\max\quad  \mu \\
\st  \quad \begin{pmatrix}
1 & \frac{\mu }{2} + \alpha  (1-\mu ) & \frac{\mu }{2}+2\alpha(1-\mu) & \frac{\mu }{2} + \alpha  (1-\mu ) & \frac{\mu }{2}+2\alpha(1-\mu) & \frac{\mu }{2} & \frac{\mu }{2} + \alpha  (1-\mu ) & \frac{\mu }{2} + \alpha  (1-\mu ) & 0 \\
  & \frac{\mu }{2} + \alpha  (1-\mu ) & 0 & \frac{\mu }{2} & \frac{\mu }{2} + \alpha  (1-\mu ) & \frac{\mu }{2} & \frac{\mu }{2} + \alpha  (1-\mu ) & \frac{\mu }{2} & 0 \\
  &  & \frac{\mu }{2}+2\alpha(1-\mu) & \frac{\mu }{2} + \alpha  (1-\mu ) & 0 & \frac{\mu }{2} & 0 & \frac{\mu }{2} + \alpha  (1-\mu ) & 0 \\
  &  &  & \frac{\mu }{2} + \alpha  (1-\mu ) & 0 & \frac{\mu }{2} & \frac{\mu }{2} & \frac{\mu }{2} + \alpha  (1-\mu ) & 0 \\
  &  &  &  & \frac{\mu }{2}+2\alpha(1-\mu) & \frac{\mu }{2} & \frac{\mu }{2} + \alpha  (1-\mu ) & 0 & 0 \\
  &  &  &  &  & \frac{\mu }{2} & \frac{\mu }{2} & \frac{\mu }{2} & 0 \\
  &  &  &  &  &  & \frac{\mu }{2} + \alpha  (1-\mu ) & \frac{\mu }{2} & 0 \\
  &  &  &  &  &  &  & \frac{\mu }{2} + \alpha  (1-\mu ) & 0 \\
  &  &  &  &  &  &  &  & 0 \\
\end{pmatrix} \ge 0
\end{gathered}
\end{equation}
We note that the only variable remaining is $\mu$. The SDP is so simple that we do not need the dual for an analytical solution. With the help of a computer algebra system, we compute the eigenvalues of the above matrix. One of them is given by
\begin{equation}
\frac{1}{76} \left(\left(4 \sqrt{5}-17\right) \mu -4 \sqrt{5}+36-\sqrt{\left(160 \sqrt{5}+1201\right) \mu ^2-16 \left(\sqrt{5}-85\right) \mu -144 \sqrt{5}+688}\right).
\end{equation}
We see that a necessary condition for it to be non-negative is that $\mu \le 5\sqrt{5}-11$. Since we know that $l_2(\mu)$ belongs to the quantum set for $\mu = 5\sqrt{5}-11$, this is the optimal value of the SDP. As a sanity check, we can substitute this value of $\mu$ in the matrix and check that it is indeed positive semidefinite.

\acknowledgments

We thank Ashutosh Rai for useful discussions. M.A. acknowledges funding from the FWF stand-alone project P 35509-N.

\bibliography{biblio}

\end{document}